\documentclass[sn-mathphys,Numbered]{sn-jnl}% Math and Physical Sciences Reference Style

\usepackage{graphicx}%
\usepackage{multirow}%
\usepackage{amsmath,amssymb,amsfonts}%
\usepackage{amsthm}%
\usepackage{mathrsfs}%
\usepackage[title]{appendix}%
\usepackage{xcolor}%
\usepackage{textcomp}%
\usepackage{manyfoot}%
\usepackage{booktabs}%
\usepackage{algorithm}%
\usepackage{algorithmicx}%
\usepackage{algpseudocode}%
\usepackage{listings}%

\def\4He{$^4$He}
\newcommand{\rhos}{\ensuremath{\rho_s}}
\newcommand{\rhon}{\ensuremath{\rho_n}}
\newcommand{\vn}{\ensuremath{v_\mathrm{n}}}
\newcommand{\vs}{\ensuremath{v_\mathrm{s}}}
\newcommand{\bv}[1]{\ensuremath{\mathbf{#1}}}

\begin{document}

\title[Article Title]{Detection of Quantized Vortices using Fourth Sound Attenuation}

\author[1]{\fnm{Filip} \sur{Novotn\'{y}}}\email{filip.novotny@mff.cuni.cz}
\author[1]{\fnm{Marek} \sur{Tal\'{i}\v{r}}}
\author[1]{\fnm{Ritesh} \sur{Dwivedi}}
\author[1]{\fnm{Šimon} \sur{Midlik}}
\author*[1]{\fnm{Emil} \sur{Varga}}\email{emil.varga@matfyz.cuni.cz}

\affil*[1]{\orgdiv{Faculty of Mathematics and Physics}, \orgname{Charles University}, \orgaddress{\street{Ke Karlovu 3}, \city{Prague}, \postcode{121 16}, \country{Czech Republic}}}

\abstract{Superfluid helium confined to nanofluidic systems is emerging as an important system for studies of two-dimensional turbulence and as the basis for novel quantum technologies. In fully enclosed nanofluidic geometries only the fourth sound can propagate, which we show can be used for probing quantized vortices pinned in well-defined slab geometry. We show that similarly to well-established second sound attenuation, fourth sound attenuation can be used to infer the number of quantized vortices in a unit area. We experimentally verify fourth sound attenuation as a probe of vortex line density by injecting a known number of quantized vortices into a nanofluidic slab using a rotating cryostat.}


\maketitle

\section{Introduction}\label{sec1}

Superfluid \4He confined to microscopic geometries, typically as thin films, is an important system for fundamental studies of, e.g., two-dimensional turbulence \cite{Gauthier, Varga2} and phase transitions \cite{Varga3, Gasparini2008} or for implementing emerging quantum technologies using electrons on helium surface \cite{Kawakami2023} or strongly non-linear confined acoustic modes \cite{Sfendla2021} as qubit platforms.

Liquid \4He becomes superfluid below 2.17~K at saturated vapour pressure. Superfluid \4He (He II) can be described as a two-component fluid as a mixture of the \emph{normal component}, which is viscous, carries the whole entropy of the system and can be treated using classical fluid mechanics, and the \emph{superfluid component}, which is inviscid and its circulation is quantized $\Gamma = n \kappa$, where $\kappa \approx $ 9.97 $\times$ 10$^{-8}$ m$^2$s$^{-1}$ is the quantum of circulation and $n$ is an integer, as first proposed by Onsager \cite{Onsager} and further developed by Feynman \cite{Feynman}. The total density of He II is then the sum of densities of both components $\rho = \rhon + \rhos$. 

The quantized circulation can take non-zero values only in multiple connected domains. In bulk He II superfluid vorticity is restricted to quantized vortices, which are topological defects with a normal core of roughly 0.1~nm diameter, each carrying a single quantum of circulation \cite{Tilley_book}. For a container of He II rotating with angular velocity $\Omega$, the superfluid will mimic solid body rotation with vorticity $\omega=2\Omega$ by becoming threaded by a regular array of quantized vortices with the average number of vortices per unit area \cite{Tilley_book}
\begin{equation}\label{eq:Feynman}
    L = \frac{2 \Omega}{\kappa}.
\end{equation}

The cores of quantized vortices scatter thermal excitations, which results in dissipative coupling between the normal and superfluid components called mutual friction \cite{Tilley_book}. Turbulence in the superfluid component takes the form of a highly disordered tangle of quantized vortices \cite{Barenghi2014}. The dynamics of quantized vortices were investigated by trapping a single quantized vortex on a microscopic wire \cite{Vinen1, Guthrie, Neumann2014}, by direct visualisation \cite{Bewley, Peretti} or, especially in turbulent cases, by attenuation of second sound \cite{Varga0}. Hall and Vinen \cite{Hall1956a, Hall1956b} have shown that the second sound (a wave in temperature or entropy density with oppositely oscillating normal and superfluid velocities) propagating perpendicularly to the cores of quantized vortices is attenuated.

In this work, we investigated the effects of quantized vortices on the attenuation of \emph{fourth} sound. In the case of the fourth sound, only the superfluid component moves and the normal component remains at rest. This can be achieved by confining He II in one of the dimensions which will result in viscous clamping of the normal fluid component to the walls. Fourth sound can be driven by, e.g., mechanically forcing the oscillatory helium flow through a porous medium \cite{Hubert}, confined channels \cite{Varga2, Varga3}, or micropores \cite{Backhaus_exp}. Due to the relative motion of normal and superfluid components in the fourth sound, the presence of quantized vortices should lead to similar damping as for the second sound. 

We demonstrate the fourth sound as a probe for investigating quantized vortices in a fully confined nanofluidic system. In analogy with second sound experiments \cite{Varga0}, we show the attenuation of the fourth sound can be used to measure the vortex line density in nanofluidic channels. We verify experimentally the obtained expression for fourth sound attenuation in a rotating cryostat.

\section{Fourth sound resonance attenuation}

\begin{figure}[h!]
    \centering
    \includegraphics[]{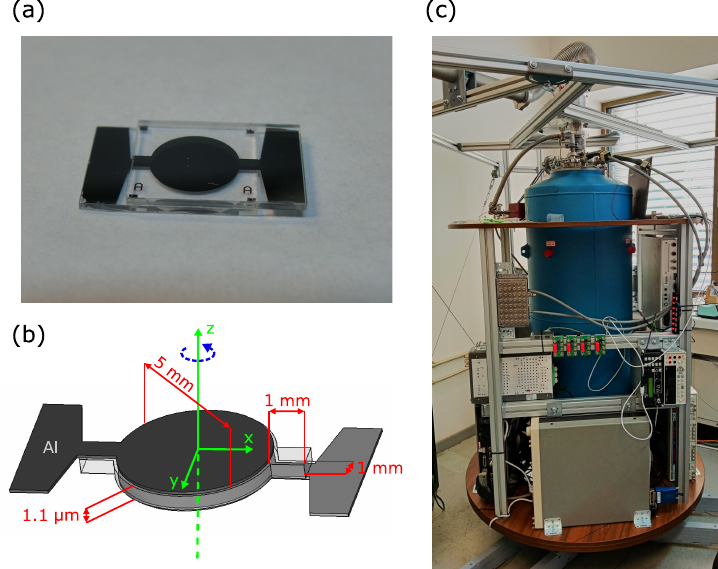}
    \caption{(a) Photo of the chip used for measurement. Two fused silica chips with etched nanofluidic channels are bonded together to create the fully enclosed nanofluidic volume. (b) Sketch of the chip with dimensions, the transparent part symbolizes the volume filled with He II. The green line is the axis of the rotation which is perpendicular to the basin and interests it in the centre. (c) Photo of the cryostat placed on the rotation platform.}
    \label{fig1}
\end{figure}

We begin by deriving the equations of motion of the fourth sound resonance in the presence of damping partly due to mutual friction. We assume that only the superfluid component moves (velocity $\bv{\vs}$) and that the normal component is at rest, $\bv{\vn}=0$. The flow will be described in the frame of the geometry used for our experiment \cite{Rojas, Souris}, see Fig.~\ref{fig1}(a,b) for a photograph and a sketch of the device. Two nanofluidic channels, with cross-section $a$ and length $l$, connect a circular volume of the same height (``basin'') with the bulk He II in which the entire nanofluidic device is submerged. The top and bottom walls of the basin are covered by aluminium, which forms a parallel plate capacitor used to drive and detect the resonance motion of the superfluid in and out of the basin. The motion of $\bv{\vs}$ will be treated in the linear approximation of small velocities and small temperature gradients, neglecting the nonlinear convective term and the thermomechanical force will be treated as a part of an intrinsic damping \cite{Souris}. The equation of motion for $\bv{\vs}$ can be written in the form
\begin{equation}
\label{eq:euler}
    \rhos \dfrac{\partial \bv{\vs}}{\partial t} = -\dfrac{\rhos}{\rho} \nabla p - \bv{f},
\end{equation}\label{Vs dynEq}
where $\rho$ and $\rho_s$ are the total and superfluid densities, respectively and $\nabla p$ is the pressure gradient along the channel, which can be simplified as $\nabla p \approx \delta P / l $ with $\delta P$ the pressure difference induced by compression of He II in the basin. Following a similar derivation presented in the supplementary material of \cite{Varga2} and \cite{Backhaus_the}, $\delta P$ as a function of a driving force $F$ can be expressed as
\begin{equation}
   \label{eq:delta-p-x}
   \delta P =  \dfrac{1}{\rho \chi V_B} (2 \rho A x - 2 a \rhos y), 
\end{equation}
where $y$ is the displacement of the fluid inside the channels, $x$ is the average displacement of the basin walls, $\chi$ the isothermal compressibility, $V_B = AD$ the volume of the basin with $A$ its area and $D$ the height of the confined volume.

The electrostatic force $F$ applied between the basin walls must be in balance with the elastic stress in the basin walls and the pressure of the helium, i.e.,
\begin{equation}
    \label{eq:force-balance}
    F = k_p x + A \delta P,
\end{equation}
where $k_p$ is the effective spring constant of the substrate. Expressing the basin compression $x$ in \eqref{eq:delta-p-x} using \eqref{eq:force-balance} we obtain
\begin{equation}
    \label{eq:deltaP}
    \delta P = \dfrac{k_p}{\rho(\chi V_B k_p + 2 A^2)} \left(  \dfrac{2 \rho A}{k_p}F - 2 a \rhos y\right).
\end{equation}
Further, we can set $\bv{\vs} = \dot y \hat{\bv x}$, where $\hat{\bv x}$ is a unit vector along the channel, under an assumption that the flow is restricted to the channels and preferentially oriented perpendicularly to the plane of the channel cross-section (i.e., flow is in-plane). Using \eqref{eq:deltaP} in \eqref{eq:euler}, we obtain 
\begin{equation}\label{eq:dynamEq}
    \rhos \ddot{y} + \dfrac{2 a \rhos^2}{\rho^2} \dfrac{k_p}{l(\chi V_B k_p + 2A^2)} y + \bv{f} \cdot \dfrac{\bv{\vs}}{\lvert \bv{\vs}\rvert} = \dfrac{\rhos}{\rho} \dfrac{2A}{l(\chi V_B k_p + 2A^2)}F.
\end{equation}

We write the friction force $\bv{f}$ per unit volume as a sum of two contributions
\begin{equation}\label{eq:totalf}
    \bv{f} = \bv f_\mathrm{i} + \bv f_\mathrm{ns}.
\end{equation}
The first term $\bv f_\mathrm{i}$ is the intrinsic friction force originating from thermal losses in the basin and residual motion of the normal fluid component, detailed discussion can be found in \cite{Souris}. Here, it is sufficient to assume this force is linearly proportional to the velocity $\bv f_\mathrm{i} = \rhos \gamma_0 \bv{\vs}$, where $\gamma_0$ is the intrinsic damping coefficient. 

The second term $\bv f_\mathrm{ns}$ is the mutual friction force, which describes the interaction of the normal fluid with quantized vortices. The force per unit length of the vortex can be written as \cite{Donnelly_QVs}
\begin{equation}\label{eq:mutualf}
    \bv f'_\mathrm{ns} = \rhos \kappa \alpha \bv s' \times [\bv s' \times (\bv{\vn} - \bv{\vs})]  + \rhos \kappa \alpha' \bv s' \times (\bv{\vn} - \bv{\vs}),
\end{equation}
where $\alpha$ and $\alpha'$ are empirical mutual friction constants \cite{Donnelly} and $\bv s$ is the spatial curve representing the quantized vortex line with $\bv s'$ the unit tangent. This can be simplified, since $\bv{\vn}$=0, and in the thin slab geometry studied here, the vortices are unlikely to form intrinsically three-dimensional structures such as loops or strongly-excited kelvin waves unless strongly forced \cite{Varga2}. Therefore, for our geometry $\bv s' \approx \pm\hat{\bv z}$ ($\hat{\bv z}$ being the unit vector perpendicular to the plane of the flow,) where the direction of the circulation around the vortex gives the sign.

Further, the second term in \eqref{eq:mutualf} is non-dissipative since it is perpendicular to $\bv{\vs}$ and can be omitted. This results in force per unit length of the vortex $\bv f'_\mathrm{ns} = \rho_s\kappa\alpha \bv\vs$ which no longer depends on the sign of the vortex. Thus, the total mutual friction force per unit volume is
\begin{equation}
    \bv{f_{ns}} =  \alpha \kappa \rhos L  \bv{\vs}  =  \rhos \gamma (L)  \bv{\vs},
\end{equation}
where $L$, the vortex line density, is the total length of vortices per unit volume, or, for the present geometry equivalently the number of vortices per unit area, and $\gamma(L) = \alpha \kappa L$ is a vortex-induced damping factor. In a rotating container, the superfluid component will mimic the solid-body rotation by creating a lattice of quantized vortices with average density $L = 2\Omega / \kappa$, where $\Omega$ is the angular velocity of the rotation.

Using the total effective friction \eqref{eq:totalf} in the equation of motion \eqref{eq:dynamEq} we get an equation of a linearly damped harmonic oscillator
\begin{equation}\label{eq:LHO}
    \ddot{y} + \tilde{\gamma} \dot{y} + \nu_0^2 y = f_0 e^{i \nu t},
\end{equation}
where $\tilde{\gamma}$ is the total damping coefficient,
\begin{equation}
    \tilde{\gamma} = \gamma_0 + \gamma (L) = \gamma_0 + \alpha \kappa L,
\end{equation}
and $\nu_0$ is the Helmholtz resonance frequency,
\begin{equation}
    \label{eq:resoanance-frequency}
    \nu_0^2 = \dfrac{2 a \rhos}{\rho^2} \dfrac{k_p}{l(\chi V_B k_p + 2A^2)}.
\end{equation}

Assuming that the electrostatic driving force $F(t) = F_0e^{i \nu t}$ the reduced driving force amplitude $f_0$ is given by
\begin{equation}
    f_0 = \dfrac{1}{\rho} \dfrac{2A}{l(\chi V_B k_p + 2 A^2) }F_0.
\end{equation}

Now we can proceed to a relation between $L$ and an attenuation of the resonance. The complex response given by \eqref{eq:LHO} to periodic driving at frequency $\nu$ is
\begin{equation}\label{eq:lorentz}
    \tilde y_0(\nu) = \dfrac{f_0}{\nu_0^2 - \nu^2  + i \nu \tilde{\gamma}},
\end{equation}
which describes the motion in-phase and $\pi/2$-shifted with the driving force detected by a lock-in amplifier. On resonance $\nu = \nu_0$ we consider two situations:
\begin{enumerate}
    \item Vortex-free state, $L = 0$, i.e., $\tilde{\gamma} = \gamma_0$.
    \begin{equation}
        \label{eq:y-vf}
        \tilde y^\mathrm{V-F}_0(\nu_0) = Y_0 = \dfrac{f_0}{i \nu_0 \gamma_0}
    \end{equation}
    \item Vortex state, $L > 0$, i.e., $\tilde{\gamma} = \gamma_0 + \alpha \kappa L$.
    \begin{equation}
        \label{eq:y-v}
        \tilde y^\mathrm{V}_0(\nu_0) = Y = \dfrac{f_0}{i \nu_0 (\gamma_0 + \alpha \kappa L)}.
    \end{equation}
\end{enumerate}
Expressing $L$ from \eqref{eq:y-vf} and \eqref{eq:y-v} we arrive at
\begin{equation}\label{eq:VLD}
    L = \dfrac{2 \pi \Delta_0}{\alpha \kappa} \left( \dfrac{Y_0}{Y} - 1\right),
\end{equation}
where $\Delta_0 = \gamma_0/2\pi$ is the vortex-free resonance width. It is thus possible to determine the number of quantized vortices in a confined system by measuring the un-attenuated and attenuated amplitude of the fourth sound resonance. The expression \eqref{eq:VLD} is similar to the expression relating second sound attenuation to vortex line density \cite{Babuin} differing only in the prefactor. A subtle difference should be noted, however, that in the form of the mutual friction \eqref{eq:mutualf}, it was assumed that the vortices move with respect to the stationary normal fluid with velocity $\bv v_s$. While the same assumption is used for the analysis of second sound attenuation \cite{Babuin}, for the case of second sound, the normal fluid also moves with respect to the vortices. This results in the temperature dependence of the prefactor in \eqref{eq:VLD} to differ as well between the fourth and the second sound, rather than the difference being a simple rescaling. It should be noted that a "vortex-free" state refers to extrinsically generated vortices, i.e. by turbulence or, in the present case, rotation. A population of pinned vortices will likely exist in the nanofluidic channel \cite{Awschalom1984}, contributing to the intrinsic damping $\gamma_0$.

\section{Experimental results}
\subsection{Experimental setup}
Fourth sound attenuation was studied using nanofluidic Helmholtz resonators (see Fig.~\ref{fig1}a,b) placed inside a rotating bath cryostat (see Fig.~\ref{fig1}c). Similar devices were used previously to study quasi-2D turbulence \cite{Varga2}, superfluid phases of $^3$He \cite{Shook2020} and finite-size effects in superfluid $^4$He \cite{Varga3}. The resonators consist of two amorphous quartz chips (7.0~mm x 9.4~mm) with etched channels bonded together to form an enclosed nanofluidic cavity with a height $D \approx 1.1$~$\mu$m. The basin radius was 5 mm and the channel size $1\times1$~mm$^2$ (see Fig.~\ref{fig1}b). The resulting resonance frequency $\nu_0$/(2$\pi$) at 1.25~K was measured $\nu_0/(2\pi)\approx 2943$~Hz. The viscous penetration depth \cite{Schmoranzer} of the normal component $\delta = \sqrt{2 \eta / \rho_n\nu} \approx$ 5.9~$\mu$m, where $\eta$ is the dynamic viscosity and $\nu$ is the angular frequency of motion, is 5-times larger than the confinement $D$ ensuring that the normal fluid is viscously clamped to the walls and the assumption of $\bv v_n= 0$ is valid.

\begin{figure}[h!]
    \centering
    \includegraphics[width = \textwidth]{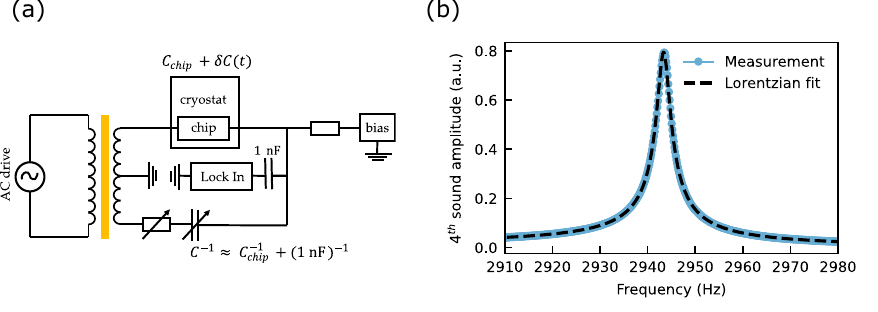}
    \caption{(a) Sketch of the detection circuit. The studied device inside the cryostat is wired as one arm of a capacitance bridge excited by a centre-tapped transformer. The bridge is balanced using a variable capacitor and resistor in the second arm. The pressure oscillation of the Helmholtz mode will induce an oscillation of the capacitance of the device, which in turn results in an oscillating current through the detector. (b) Typical amplitude response curve. The dashed line shows the fit to \eqref{eq:lorentz} with the addition of a linear background.}
    \label{fig10}
\end{figure}

To induce and detect the motion of the fluid, two aluminium electrodes were evaporated on the top and bottom walls of the basin, which form a parallel plate capacitor of capacitance $C_0$. By applying an AC voltage $u(t)$ to the basin electrodes biased by DC voltage $U_B = 10$~V, electrostatic force $F(t) = C_0U_Bu(t)/D$ (neglecting off-resonant terms) deforms the basin walls and thus creates a pressure difference between the basin and the surrounding bulk He II. The resonant motion of the fluid results in pressure oscillation in the basin, which deforms its walls and thus the capacitance of the device, which when biased by DC voltage yields the measured AC current. To suppress the background signal due to the drive, the device was wired in a bridge circuit, see Fig.~\ref{fig10}(a) for the sketch of the circuit and refs.~\cite{Shook2020, Varga2, Varga2021} for further details.

The device was placed on the axis of the rotating helium cryostat with the plane of the flow perpendicular to the rotation axis. The rotating platform could reach angular velocities up to 180 deg s$^{-1}$ in the present experiment. The helium bath was cooled to 1.250~K (measured by a semiconductor resistive thermometer calibrated against saturated vapour pressure) by pumping on the helium vapours and the temperature was stabilised to approximately 1~mK with a resistive heater controlled by a PID loop.

\subsection{Results}
\begin{figure}[h!]
    \centering
    \includegraphics[width = \textwidth]{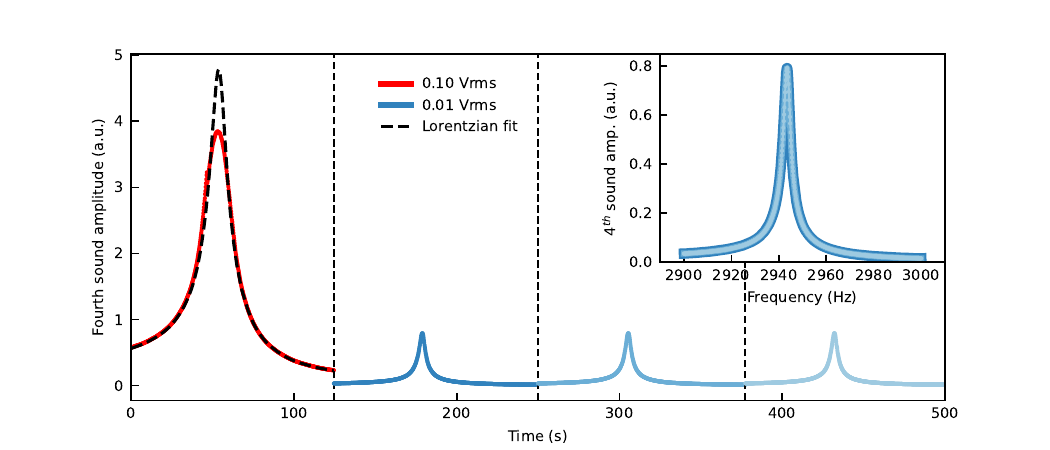}
    \caption{Measurement sequence showing the recorded 4th sound amplitude as a function of time while sweeping the drive frequency across the resonance. The first peak, red, is driven into the nonlinear regime and serves to ``anneal'' the vortex configuration (see text). The jump on the left slope of the peak marks the discontinuous transition turbulence \cite{Varga2}. The Lorentzian fit indicates the linear behaviour of the annealing peak tails. The following three peaks, blue, are driven in the linear regime and serve as the detection peaks. The inset shows detection peaks in detail against the frequency of the driving signal, indicating the consistency in the peak's shape. The legend shows the resonance excitation amplitude.}
    \label{fig2}
\end{figure}

The attenuation of the 4th sound was characterized using a measurement sequence shown in Fig.~\ref{fig2}. Before the measurement, the cryostat was left to rotate at the target angular velocity for at least one minute. With steady rotation, the Helmholtz mode was driven sufficiently strongly (the first red curve in Fig.~\ref{fig2}) to show nonlinear dissipation due to turbulence \cite{Varga2}. Next, the resonant response was measured with a sufficiently weak drive to remain in the linear regime (second to fourth blue peaks in Fig.~\ref{fig2}). The peak amplitude entering \eqref{eq:VLD} was evaluated using the peaks in the linear regime.

Without the first strongly--driven non-linear peak the observed attenuation in the linear regime did not show any well-defined dependence on the rotation speed even when left under steady rotation for up to approximately 1 hour. The vortices likely pinned in the device appear to be in a metastable state with very slow relaxation to the equilibrium value given by the angular velocity $\Omega$ and \eqref{eq:Feynman}, which can be annealed by sufficiently strong flow. Strong turbulence induced by the annealing peak is short-lived in comparison with the time of the frequency sqweep, since the peak tails follow the linear behaviour as it is shown by the Lonrentzian fit, see Fig.~\ref{fig2} and we do not observe any time dependence among the three linear measurement sweeps (i.e., there is no observable slow decay). We recall that in the derivation of \eqref{eq:VLD}, it was assumed that the vortices move with the superflow. Especially at low probe amplitudes, pinning will likely play a role in the sensitivity of the fourth sound to the presence of vortices. The nature of the metastable states and their annealing and the role of pinning will be the subject of future experiments.

\begin{figure}[h!]
    \centering
    \includegraphics[width = \textwidth]{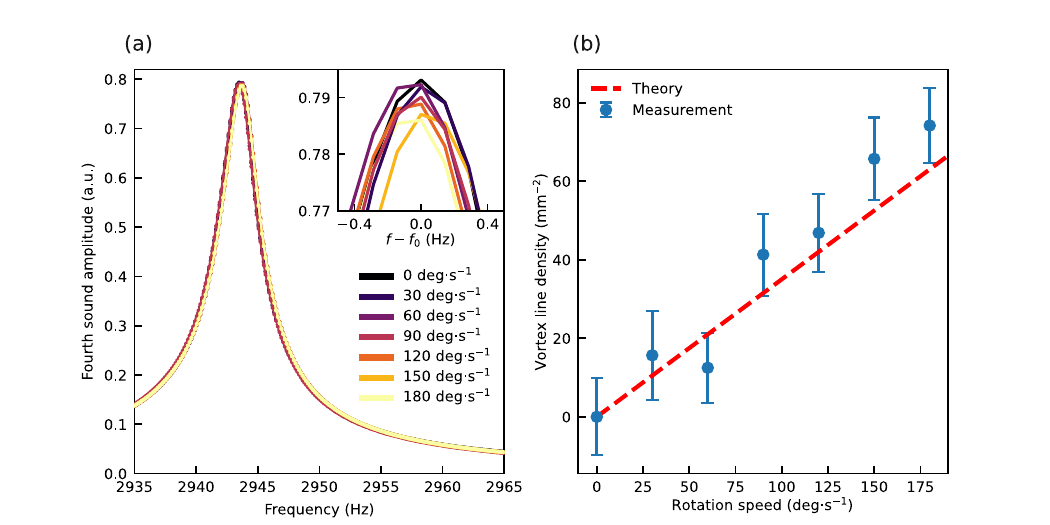}
    \caption{(a) Typical resonant amplitude response in steady state and under the rotation, used for the calculation of $L$, against the frequency of the driving signal. Shown are the averages of several resonance curves measured at each angular velocity. In the inset, the detail of the resonance curves maximum is shown to emphasize the decrease of the amplitude used to calculate $L$ using \eqref{eq:VLD}. The maxima of peaks are shifted to zero by subtracting the resonance frequency of the given peak $f_0$. (b) Vortex line density $L$ against the angular velocity of the rotating cryostat. Blue points correspond to the measured values by the attenuation of the fourth sound, red dashed line is the theoretical linear dependence given by the Feynman rule \eqref{eq:Feynman}. The error bars indicate the variability of the measured amplitude on resonance (details in the text).}
    \label{fig3}
\end{figure}

We calculated values of $L$ using \eqref{eq:VLD}, where amplitudes $Y_0$, $Y$ and the width $\Delta_0$ were obtained by a fit of the data measured at 10~mVrms excitation to \eqref{eq:lorentz} (with additional linear background), for a typical fit see the Fig.~\ref{fig10}(b). The parameters of the unattenuated peak $Y_0$ and $\Delta_0$ were measured at zero rotation speed and averaged for calculation of $L$. Set of averaged peaks is shown in the Fig.~\ref{fig3}(a). From the inset, one can see, that the change in the height of peaks between the steady and the rotation state is relatively small at the maximum approximately 1\% depending on the rotation speed. To detect such small changes, a fourth sound resonance with a sufficiently high quality factor $Q = \nu_0/(2\pi\Delta_0)$ is needed. In our case, the quality factor at 1.25 K (close to the base temperature of the cryostat) was $Q \approx 1060$. Further, peaks under rotation were slightly shifted (tenths of Hz) to higher frequencies than those in the stationary case. Similar, stronger, shifts of fourth sound resonance to lower frequencies were observed in the experiment with Al$_2$O$_3$ powder under rotation \cite{Hubert} and speculatively explained by an effective lengthening of the fourth sound resonance length due to quantized vortices. In our case, the shift is likely due to a slight decrease in pressure at the centre of the rotation, since the entire helium bath is set to rotate. A decrease of pressure leads to an increase in the superfluid density $\rho_s$ \cite{Brooks1977}, which would lead to an increase in the resonance frequency according to \eqref{eq:resoanance-frequency}.

The mutual friction parameter $\alpha$ at 1.250~K was linearly extrapolated from values reported by Donnelly and Barenghi \cite{Donnelly} in the range from 1.300~K to 1.375~K. At each rotation speed, we obtained at least 2 sequences shown in Fig.~\ref{fig2}, i.e., at least six resonance curves for each rotation velocity. Averaged values of $L$ are shown in Fig.~\ref{fig3}(b) against the speed of the rotation in comparison with the theoretical relation \eqref{eq:Feynman}. 

The measured vortex line density $L$ against the angular velocity of the cryostat is in reasonable agreement with the theory following the predicted linear trend. Values of $L$ are in tens of vortices per mm$^2$. The systematic underestimation of the experimental data by the theory is likely either due to the fourth sound resonance being already slightly non-linear or due to the uncertain value of the mutual friction constant $\alpha$ entering \eqref{eq:VLD} which needed to be extrapolated from higher temperatures. The outlier at 60 deg s$^{-1}$ was affected by a slight temperature instability. The peak parameters entering \eqref{eq:VLD} were obtained from a fit of measured resonance curves to \eqref{eq:lorentz}, which yielded the parameters $X$ and the estimate of their variance $\sigma_{X,\mathrm{fit}}^2$ ($X$ is $f$ or $\gamma$ in \eqref{eq:lorentz}, from which $Y$ and $\Delta_0$ are calculated). The resonance curves were measured several times under nominally identical conditions, yielding a set of fit parameters $X_i$ and $\sigma_{X_i,\mathrm{fit}}^2$. The value of $X$ entering the calculation of $L$ was determined as the average of $X_i$ weighted by the inverse variance $1/\sigma_{X_i,\mathrm{fit}}^2$ and the final error of the parameter was calculated as $\sigma_X^2 = \mathrm{Var}(X) + \sigma_\mathrm{fit}^2$, where the variance $\mathrm{Var}(X)$ is calculated form the ensemble $\{X_i\}$ and $\sigma_\mathrm{fit}^2 = (\sum_i \sigma_{X_i,\mathrm{fit}}^{-2})^{-1}$. The error bars of $L$ in Fig.~\ref{fig3}b were then calculated using errors of parameters $\sigma_X$.

The error bars are conservatively increased by the statistical error of the fit parameters, however they do not appear to fully explain the slight systematic descrepancy from the theoretically predicted trend. Considering that the superflow is mainly localised in the channels \cite{Rojas, Souris, Varga2}, with an area of 1~mm$^2$, the measured values of $L$ indicate that the measurement is sensitive to as few as 20 vortices being additionally present due to rotation. However, it is necessary to emphasize that the fourth sound resonance attenuation can only provide information about the average value of vortex density in the studied area and not the spatial distribution. Nevertheless, in the current experiment, we were able to resolve $L$ with the accuracy of approximately 5 vortices per mm$^{2}$. 

\section{Conclusion}
We measured the vortex line density $L$ in quasi 2D superfluid helium using the fourth sound attenuation with quantized vortices generated independently on the fourth sound resonance by the rotation of the cryostat. Using methods similar to second sound attenuation \cite{Varga0} $L$ is calculated from the attenuated amplitude of the fourth sound resonance. We showed, that measured values follow the trend given by the Feynman rule \eqref{eq:Feynman}. With further improvement of the system stability (i.e., temperature and mechanical vibration) and detection peak sensitivity, which can be improved by increasing $Q$ either by lower temperatures or optimized flow geometry, single-vortex sensitivity appears within reach. The attenuation of the fourth sound in engineered nanofluidic structures thus presents a promising platform for studying vortex dynamics in quasi-2D superfluid helium ranging from few--vortex clusters to strongly turbulent flows.

\subsubsection*{Acknowledgements}
The work was supported by the Charles University under PRIMUS/23/SCI/017. CzechNanoLab projects LM2023051 and LNSM-LNSpin funded by MEYS CR are gratefully acknowledged for the financial support of the sample fabrication at CEITEC Nano Research Infrastructure and LNSM at FZU AV\v{C}R and we are grateful to K. Olejník for providing access to the FZU cleanrooms. We are also grateful to L. Skrbek, D. Schmoranzer, P. Urban, I. Vl\v{c}ek and M. Zoba\v{c} for their help with the construction of the rotating platform.

\section*{Declarations}
The authors have no competing interests to declare.



\end{document}